\documentclass[letter,12pt]{article}

\usepackage{pstricks}
\usepackage{epsfig}
\usepackage{amsmath}
\usepackage{bm}
\usepackage{array}
\usepackage{subfig}
\usepackage{graphicx}
\usepackage{cite}
\usepackage{amssymb}

\def\({\left(}
\def\){\right)}

\begin{document}
\setlength{\baselineskip}{18pt}
\vspace{-3cm}

\begin{flushright}
OSU-HEP-12-08
\end{flushright}

\renewcommand{\thefootnote}{\fnsymbol{footnote}}

\begin{center}
{\Large\bf \boldmath{$B-L$} Violating Proton Decay Modes and  \\[0.1in]
New Baryogenesis Scenario in \boldmath{$SO(10)$}}\\
\end{center}

\vspace{0.5cm}
\begin{center}
{ \bf {}~K.S. Babu}$^a$\footnote{Email:
babu@okstate.edu} and {\bf R.N. Mohapatra}$^b$\footnote{Email: rmohapat@umd.edu}
\vspace{0.5cm}

{\em $^a$Department of Physics, Oklahoma State University,
Stillwater, OK 74078, USA }
\vspace*{0.3cm}

{\em $^b$Maryland Center for Fundamental Physics, Department of Physics,\\ University of Maryland,
College Park, MD 20742, USA }

\end{center}

\begin{abstract}
\setlength{\baselineskip}{18pt}

We show that grand unified theories based on $SO(10)$ generate quite naturally
baryon number violating dimension seven operators that violate $(B-L)$,
and lead to novel nucleon decay modes such as $n \rightarrow e^-K^+, e^- \pi^+$ and $p \rightarrow \nu \pi^+$.
We find that in two--step breaking schemes of non-supersymmetric $SO(10)$, the partial lifetimes
for these modes can be within reach of experiments.
The interactions responsible for these decay modes also provide a new way to understand the origin of matter
in the universe via the decays of GUT scale scalar bosons of $SO(10)$.
Their $(B-L)$--violating nature guarantees that the GUT scale induced baryon asymmetry
is not washed out by the electroweak sphaleron interactions.
In minimal $SO(10)$ models this asymmetry is closely
tied to the masses of quarks, leptons and the neutrinos.
\end{abstract}

\newpage
\renewcommand{\thefootnote}{\arabic{footnote}}
\setcounter{footnote}{0}

Baryon number violation is a very sensitive probe of physics beyond the Standard Model (SM).
Interactions which violate baryon number ($B$) are not present in the renormalizable part of the SM Lagrangian,
but can arise through effective higher dimensional operators.
The leading $B$ violating operators \cite{weinberg} have dimension six and are hence
suppressed by two powers of an inverse mass scale. These operators arise naturally when SM is embedded in grand
unified theories (GUTs) such as $SU(5)$ and $SO(10)$. They lead to nucleon decay modes such as $p \rightarrow e^+ \pi^0$ and
$p \rightarrow \overline{\nu} K^+$, which conserve baryon number minus lepton number
$(B-L)$ symmetry. Experimental searches to date have primarily focussed on these modes with the latest limits on proton lifetime
constraining the masses of the heavy mediators to be larger than about $10^{15}$ GeV. This is
in accord with the scale inferred from the unification of gauge couplings.

Going beyond the $d=6$ baryon number violating operators, the
next--to--leading ones have $d=7$, and obey the selection rule
$\Delta (B-L) = -2$ for nucleon decay \cite{weinberg2}.  These operators
lead to novel nucleon decay modes such as $n \rightarrow e^- K^+, e^- \pi^+$, and $p \rightarrow \nu \pi^+$,
which have received less attention.
In this Letter we show that these $d=7$ operators
arise naturally in unified theories based on $SO(10)$, upon the spontaneous breaking of $(B-L)$, which is part of the gauge symmetry.
In particular, we find that in non--supersymmetric $SO(10)$ models with an intermediate scale so that gauge couplings unify,
the partial lifetime to these decay modes
can be within reach of ongoing and proposed experiments.
Furthermore, we show that these new modes provides a novel way to understand the origin of matter in the universe.
This mechanism relies on the fact that, owing to their $(B-L)$ breaking nature, a GUT scale induced baryon asymmetry
would not be affected by the electroweak sphalerons \cite{kuzmin} and would survive down to low temperatures.
Observed baryon number of the universe then carries the direct imprint of GUT scale physics.
This is unlike the $(B-L)$--preserving baryon asymmetry induced in the decays of GUT
mass particles such as in $SU(5)$, which is however washed out by the sphaleron interactions, leaving no trace of GUT physics.
We show that in minimal $SO(10)$ models \cite{babu} which have been highly successful in predicting large
neutrino oscillation angles, including a relatively
large value of $\sin^22\theta_{13}\simeq (0.085-0.095)$, consistent with recent results \cite{theta13}, the baryon
asymmetry of the right magnitude is generated by the new $(B-L)$--violating mechanism.
The results of this paper should provide motivations to search for $(B-L)$--violating semi-leptonic decay modes of the nucleon in the
ongoing and the next round of searches. Their observation would furnish evidence against the simple one--step breaking of GUT symmetry,
and could also resolve the mystery behind the origin of matter in the universe.

We start by writing down the $d=7$ $B$--violating effective operators in the SM \cite{weinberg2} in the standard notation for fermion fields:
\begin{eqnarray}
{\cal \tilde{O}}_1 &=& (d^c u^c)^* (d^c L_i)^* H^*_j \epsilon_{ij},~~~~~~{\cal \tilde{O}}_2 = (d^c d^c)^* (u^c L_i)^* H^*_j \epsilon_{ij}, \nonumber \\
{\cal \tilde{O}}_3 &=& (Q_i Q_j)(d^c L_k)^*H^*_l \epsilon_{ij} \epsilon_{kl},~~
{\cal \tilde{O}}_4 = (Q_i Q_j) (d^c L_k)^*H^*_l (\vec{\tau} \epsilon)_{ij}\cdot (\vec{\tau}\epsilon)_{kl}, \nonumber \\
{\cal \tilde{O}}_5 &=& (Q_i e^c) (d^c d^c)^*H^*_i,~~~~~~~~~~
{\cal \tilde{O}}_6 = (d^c d^c)^*(d^c L_i)^* H_i, \nonumber \\
{\cal \tilde{O}}_7 &=& (d^c D_\mu d^c)^*(\overline{L}_i \gamma^\mu Q_i),~~~~~~{\cal \tilde{O}}_8 = (d^c D_\mu L_i)^*(\overline{d^c} \gamma^\mu Q_i), \nonumber \\
{\cal \tilde{O}}_9 &=& (d^c D_\mu d^c)^* (\overline{d^c} \gamma^\mu e^c)~.
\label{dim7}
\end{eqnarray}
Here $D_\mu$ stands for the covariant derivative, and $H(1,2,1/2)$ is the Higgs doublet.
These operators obey $(B-L) = +2$ selection rule and mediate nucleon decays of the type $n \rightarrow e^- K^+,
e^-\pi^+$ and $p \rightarrow \nu \pi^+$.  We first show how these operators arise naturally in $SO(10)$ theories \cite{bm2012}
when $(B-L)$ symmetry contained in it is broken. This breaking may occur at the GUT scale as in models with supersymmetry,
or at an intermediate scale $M_I$ below the GUT scale, as in non--supersymmetrci $SO(10)$, which requires such a scale
to be compatible with gauge coupling unification.
To see the origin of Eq. (\ref{dim7}) in $SO(10)$ via scalar boson exchange, we write down the Yukawa couplings in the most
general setup.  Noting that the fermion bilinears contain $16 \cdot 16 = 10_s + 120_a + 126_s$,  the Yukawa couplings are \cite{aulakh1}:
\begin{eqnarray}
{\cal L}(16_i 16_j 10_H) &=& h_{ij} \left[ (u^c_i Q_j + \nu^c_i L_j)\, h - (d^c_i Q_j + e^c_i L_j)\, \overline{h} +
\left( \frac{\epsilon}{2} Q_i Q_j + u^c_i e^c_j - d^c_i \nu^c_j \right)\omega  \right. \nonumber \\
&& \left. + \left(\epsilon u^c_i d^c_j + Q_i L_j  \right)\omega^c \right],
\label{Yuk10}
\end{eqnarray}
\begin{eqnarray}
{\cal L}(16_i 16_j \overline{126}_H) &=& f_{ij} \left[(u^c_iQ_j - 3 \nu^c_i L_j)\,h - (d^c_iQ_j-3 e^c_iL_j)\,\overline{h} \right. \nonumber \\
&& \left.  + \sqrt{3}i\left(\frac{\epsilon}{2} Q_i Q_j - u^c_i e^c_j+ \nu^c_i d^c_j\right) \omega_1
+ \sqrt{3} i (Q_i L_j - \epsilon u^c_i d^c_j)\,\omega_1^c \right. \nonumber \\
&& \left.  + \sqrt{6}(d^c_i \nu^c_j + u^c_i e^c_j)\, \omega_2
+  2 \sqrt{3} i\, d^c_i\, L_j\, \rho  - 2 \sqrt{3} i\, \nu^c_i\, Q_j\, \overline{\rho} + 2 \sqrt{3} \, u^c_i \,\nu^c_j\,\eta \right. \nonumber \\
&& \left.- 2 \sqrt{3} i\, u_i^c \, L_j \, \chi + 2 \sqrt{3} i\,  e_i^c \,Q_j \,\overline{\chi} -2 \sqrt{3}\, d_i^c\, e_j^c\, \delta +
\sqrt{6} i \,Q_i\, L_j \,\overline{\Phi}
+ ....\right],
\label{Yuk126}
\end{eqnarray}
\begin{eqnarray}
{\cal L}(16_i 16_j 120_H) &=& g_{ij}\left[(d_i Q^j+e^c_iL_j)\,\overline{h}_1 - (u^c_i Q_j + \nu^c_i L_j)\, h_1 - \sqrt{2} Q_i L_j\, \omega_1^c \right. \nonumber \\
&& \left. - \sqrt{2} (u^c_i e^c_j - d^c_i \nu^c_j)\,\omega_1 -\frac{i}{\sqrt{3}}(d^c_i Q_j-3 e^c_iL_j)\, \overline{h}_2 + \frac{i}{\sqrt{3}}
(u^c_i Q_j-3 \nu^c_i L_j)\, h_2 \right. \nonumber \\
&& \left. -2 e^c_i Q_j \,\overline{\chi}+ 2 \nu^c_i Q_j\, \overline{\rho}-2 d^c_iL_j \, \rho + 2 u^c_i L_j \, \chi \right. \nonumber \\
&& \left. - i\, \epsilon\, d^c_i d^c_j \, \overline{\eta}+ 2 \,i\, u^c_i \nu^c_j\, \eta + \sqrt{2}\, i\, \epsilon\, d^c_i u^c_j \, \omega_2^c + \sqrt{2}\, i\,
(d^c_i \nu^c_j-e^c_i u^c_j)\, \omega_2 \right. \nonumber \\
&& \left. -\frac{\epsilon}{\sqrt{2}}Q_i Q_j \Phi - \sqrt{2}\, Q_i L_j \overline{\Phi}  -2 \,i\, d_i^c\,  e^c_j\, \delta + i\, \epsilon\,
 u_i^c\, u_j^c\, \overline{\delta} + ...\right],
\label{Yuk120}
\end{eqnarray}
with $h$ and $f$ being symmetric and $g$ being anti-symmetric in flavor indices $i,j$.
%These terms are written in terms of the SM decomposition of the sub-multiplets.
The $SU(3)_C \times SU(2)_L \times U(1)_Y$ quantum numbers of the various sub-multiplets in Eqs. (\ref{Yuk10})-(\ref{Yuk120})
are: $h(1,2,+1/2)$, $\overline{h}(1,2,-1/2)$, $\omega(3,1,-1/3)$, $\omega^c(\overline{3}, 1, 1/3)$,
$\rho(3,2,1/6)$, $\overline{\rho}(\overline{3},2,-1/6)$, $\eta(3,1,2/3),\overline{\eta}(\overline{3},1,-2/3),$
$\Phi(3,3,-1/3)$, $\overline{\Phi}(\overline{3},3,1/3)$, $\chi(3,2,7/6),\overline{\chi}(\overline{3},2,-7/6),$
$\delta(3,1,-4/3)$, and $\overline{\delta}(\overline{3},1,4/3)$.

\begin{figure}[h]
\centering
	\includegraphics[scale=0.5]{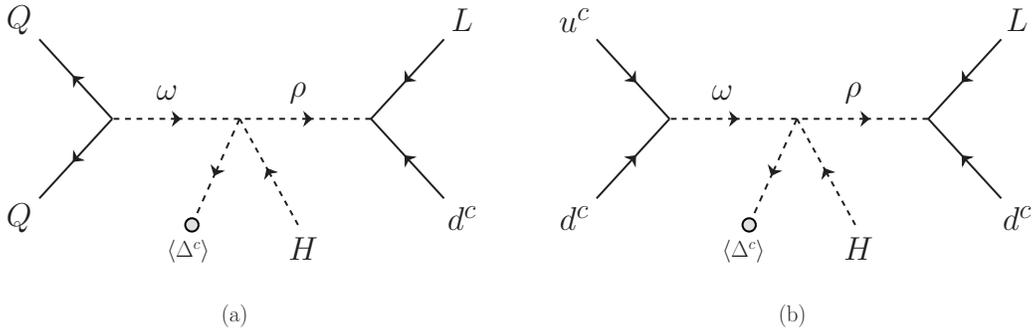}
	\caption{Effective baryon number violating $d=7$ operators induced by the symmetric Yukawa couplings of $10_H$ and $\overline{126}_H$ of $SO(10)$.}
	\label{sym}
\end{figure}

The $(B-L)$ generator of $SO(10)$ is broken by the VEV of the SM singlet field $\Delta^c$ in $\overline{126}_H$, which has $(B-L) = -2$.
This VEV supplies large Majorana masses for the right--handed neutrinos through the coupling $f_{ij}\sqrt{6} \nu^c_i \nu^c_j \Delta^c$.
It also generates trilinear scalar couplings of the type $\rho^* \omega H$, $\eta^* \rho H$, $\rho^* \Phi H$ and $\chi^* \eta H$, thereby
inducing the $d=7$ baryon number violating operators of Eq. (\ref{dim7}), via the Yukawa couplings of Eqs. (\ref{Yuk10})-(\ref{Yuk120}).
The flavor symmetric Yukawa couplings of Eqs. (\ref{Yuk10})-(\ref{Yuk126}) generate the operators ${\cal \tilde{O}}_3$ and
${\cal \tilde{O}}_1$ through the diagrams shown in Fig. \ref{sym}.

The trilinear couplings in Fig. \ref{sym} have different sources in $SO(10)$.  The quartic coupling $(126)^4$, which is invariant,
contains the term $(2,2,15) \cdot (2,2,15) \cdot (1,1,6) \cdot (1,3,\overline{10})$ under $SU(2)_L \times SU(2)_R \times SU(4)_C$
subgroup. The $\rho^*(\overline{3},2,-1/6) \subset(2,2,15)$, $H(1,2,1/2) \subset(2,2,15)$,
while $\omega(3,1,-1/3) \subset (1,1,6)$, and this coupling would contain
the term $\rho^* \omega H \overline{\Delta^c}$.  The three non-trivial invariants of the type $(126)^2 \cdot (126^*)^2$ also contain
this trilinear term.  Similarly, the three quartic couplings $(120)^2 \cdot (126)^2$
would generate the trilinear terms $\rho^* \eta H^*$, $\chi^* \eta H$, and $\Phi^* \rho H^*$ vertices, and along with Eq. (\ref{Yuk120})
would induce the remaining non--derivative $d=7$ operators of Eq. (\ref{dim7}).  For a detailed discussion see Ref. \cite{bm2012}.

Such trilinear couplings are also present when the $126_H$ is replaced by a $16_H$
albeit in
a slightly different way.   The $16_H$ contains a SM singlet filed
with $B-L = +1$ which acquires a GUT scale VEV, and a $\overline{h}(1,2,-1/2)$ field with $B-L = -1$.
The trilinear scalar couplings of the type $16_H 16_H 10_H$ and $\overline{16}_H \overline{16}_H 10_H$
would mix the $B-L = 0$ Higgs doublet $h(1,2,1/2)$  from the $10_H$ and the
$h(1,2,1/2)$ Higgs from the
$\overline{16}_H$ which has $B-L = +1$.  The light SM Higgs doublet then would have no definite $B-L$ quantum
number.  The $(1,2,4)$ component of $\overline{16}_H$ contains the field $\rho^*(\overline{3},2,-1/6)$, and the $(2,1,\overline{4})$ of $\overline{16}_H$  contains $\omega(3, 1, -1/3)$, and thus the coupling $\rho^* \omega H$ is generated via the $\overline{16}_H
\overline{16}_H 10_H$ coupling.

The $d=7$ operators of Eq. (\ref{dim7}) can also arise by integrating out
the vector gauge boson $V_Q(3,2,1/6)$ and  $V_{u^c}((\overline{3},1,-2/3)$ of $SO(10)$, which lie outside of $SU(5)$ \cite{bm2012}.
The covariant derivative for the $126_H$ would contain the term $V_Q V_{u^c} H (\Delta^c)^\dagger$
which generates the $d=7$ operators.  When $16_H$ is used instead of the $126_H$, the
covariant derivative would contain a similar term, but now with $(B-L) = +1$ and $-1$ for $H$ and $\Delta^c \subset 16_H$
respectively.

\begin{figure}[h]
\centering
\includegraphics[scale=0.73]{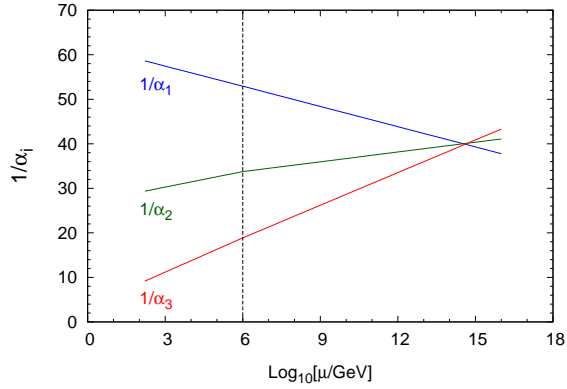}
	\caption{Unification of the three SM gauge couplings obtained with a light
$\rho(3,2,1/6)$ and two $(1,3,0)$ scalar multiplets at $M_I= 10^6$ GeV.
  }
	\label{unif}
\end{figure}

{\it Partial lifetime for $(B-L)$ violating nucleon decay:}
The diagrams of Fig. \ref{sym} lead to the following estimate for $n \rightarrow e^-\pi^+$  lifetime:\footnote{$(B+L)$--preserving nucleon decay has been studied
in the context of $R$--parity breaking SUSY in Ref. \cite{vissani}.}
\begin{equation}
\Gamma (n \rightarrow e^-\pi^+)^{\rm Fig. \ref{sym}} \approx \frac{|Y^*_{QQ\omega} Y_{L d^c \rho}|^2}{64 \pi}(1+D+F)^2 \frac{\beta_H^2 m_p}{f_\pi^2}\left(\frac{\lambda v v_R}{M^2_\rho} \right)^2
\frac{1}{M_\omega^4}~.
\label{tau}
\end{equation}
Here we have defined the Yukawa couplings of $\omega$ and $\rho$ fields appearing in Fig. \ref{sym} to be
$Y^*_{QQ\omega}$ and  $Y_{L d^c \rho}$.
The factors $D$ and $F$ are chiral Lagrangian factors, $D \simeq 0.8$
and $F \simeq 0.47$.  $\beta_H \simeq 0.012~{\rm GeV}^3$ is the nucleon decay matrix element,
$v_R \equiv \left\langle \Delta^c\right\rangle$, and $v\equiv \left \langle H^0 \right \rangle = 174$ GeV.
We have defined the trilinear coupling of Fig. \ref{sym} to have a coefficient $\lambda v_R$.
The mass of $\omega(3,1,-1/3)$ is constrained to be relatively large, as it
mediates $d=6$ nucleon decay.  For $Y\approx 10^{-3}$, $M_\omega > 10^{11}$ GeV
must be met from the $d=6$ decays.  As an illustration, choosing $Y_{QQ\omega}=Y_{L d^c \rho} =  10^{-3}$,
$M_\omega = 10^{11}$ GeV, $M_\rho = 10^8$ GeV, and $\lambda v_R = 10^{11}$ GeV in Eq. (\ref{tau}),
we find $\tau_n \approx  3 \times 10^{33}$ yrs.  Such a spectrum is motivated by
the intermediate symmetry $SU(2)_L \times SU(2)_R \times SU(4)_C$, which is found  to be
realized at $M_I \approx 10^{11}$ GeV from gauge coupling unification \cite{chang}.
As a second example, take $M_\rho = 10^6$ GeV, $M_\omega = 10^{16}$ GeV, $\lambda v_R = 10^{16}$ GeV, $Y_{QQ\omega}=Y_{L d^c \rho} = 3 \times 10^{-3}$. This choice of spectrum leads to $\tau_n \approx 4 \times  10^{33}$ yrs. This spectrum can arise as follows.
Suppose the $\rho(3,2,1/6)$ particle, along with a pair of $(1,3,0)$ scalar particles
(contained in the $45_H$, $54_H$ or $210_H$ needed for symmetry breaking) survive down to $M_I = 10^6$ GeV.  The SM gauge couplings
are found to unify at a scale $M_X \approx 10^{15}$ GeV in this case, as shown in Fig. \ref{unif}.
This scenario would predict observable rates for both the
$(B-L)$--conserving and $(B-L)$--violating nucleon decay modes. Analogous results are obtained from the exchange of $\eta(3,1,2/3)-\rho(3,2,1/6)$
scalar bosons from $120_H$. Since these particles do not induce $d=6$ baryon number violation, they can both have mass of order $M_I$, which
would enhance the nucleon decay rate.

While the $\Delta B = -2$ nucleon lifetime is in the experimentally accessible range for reasonable choice of
parameters as shown, it is quite sensitive to the precise values of the intermediate scalar masses.
For example, a factor of 3 increase in $M_\rho$ and $M_\omega$ will increase the lifetime by a factor of $10^4$.
Not finding these modes will not exclude this class of $SO(10)$ models, but a discovery of the
$\Delta B = -2$ nucleon decay mode would lend strong support to a new mechanism of baryogenesis
via $(B-L)$--violating decays of scalars, to which we now turn.

\begin{figure}[h]
\centering
	\includegraphics[scale=0.5]{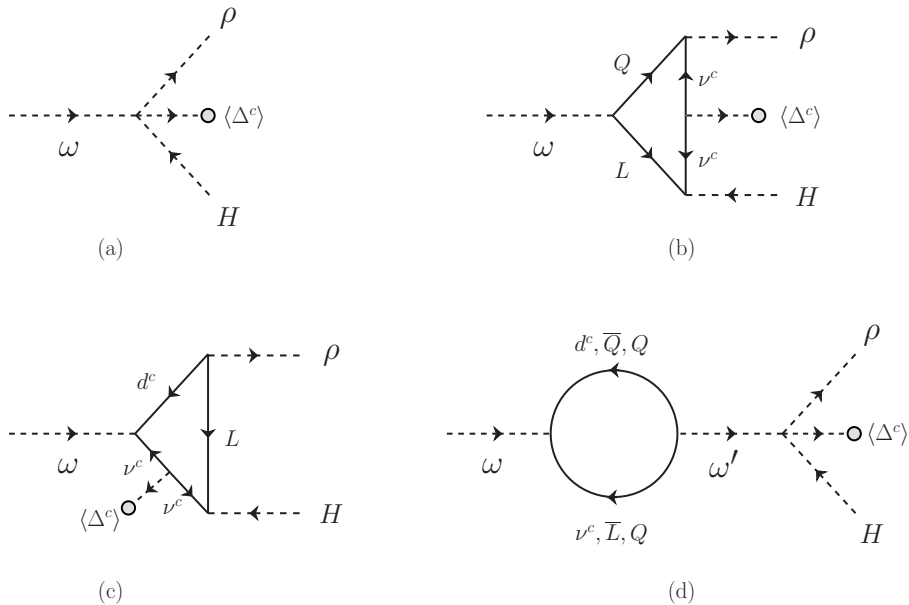}
	\caption{Tree--level diagram and one--loop corrections responsible for generating $(B-L)$ asymmetry in $\omega$ decay.}
	\label{baryo1}
\end{figure}

{\it New baryogenesis scenario at the GUT epoch:}
 We now present a new baryogenesis scenario at the GUT epoch, using the $(B-L)$--violating decay of the scalar $\omega(3,1,-1/3)$ with a GUT scale mass.  The magnitude of the asymmetry is directly linked to the neutrino masses, since the Yukawa couplings that induce the asymmetry
 are the same couplings that are involved in neutrino mass generation.
($B-L$ asymmetry in decays of specific heavy particles has recently been discussed in Ref. \cite{maekawa}.)
To be concrete, we shall work in the framework of non--supersymmetric $SO(10)$, although our results would hold for SUSY $SO(10)$
as well, with some minor modifications.    The couplings of Eq. (\ref{Yuk10})-(\ref{Yuk120}) imply that $\omega$ has two--body decays into fermions of the type
$\omega \rightarrow \overline{Q}\, \overline{Q},\,\overline{u^c}\,\overline{e^c},\,\overline{\nu^c}\, \overline{d^c},\,u^c\, d^c,\, Q\,L$.
These decays preserve $(B-L)$, as can be seen by assigning $(B-L)(\omega) = -2/3$.  Now, $\omega$ also has a two--body scalar decay,
$\omega \rightarrow \rho H^*$ as shown in Fig. \ref{baryo1} (a), which uses the $(B-L)$ breaking VEV of $\Delta^c$.
The scalar field $\rho$ has two--body fermionic decays of the type
$\rho \rightarrow \overline{L}\, \overline{d^c},\,\nu^c \,Q$ (the latter if kinematically allowed), which define $(B-L)$ charge of
$\rho$ to be $+4/3$.  Thus the decay $\omega \rightarrow \rho H^*$ would violate $(B-L)$ by $-2$.

Let the branching ratio for $\omega \rightarrow \rho H^*$ be $r$ which produces a net $(B-L)$ number of $4/3$, and that for $\omega^* \rightarrow \rho^* H$ be $\overline{r}$, with net $(B-L) = -4/3$.  The branching ratio for the two--fermion decays
$\omega \rightarrow ff$ is then $(1-r)$ which has  $(B-L) = -2/3$, and that
for $\omega^* \rightarrow \overline{f}\, \overline{f}$ is $(1-\overline{r})$ which has $(B-L) = 2/3$.
Thus in the decay of a $\omega + \omega^*$ pair, a net $(B-L)$ number, defined as $\epsilon_{B-L}$, is induced, with
$
\epsilon_{B-L}  \equiv  (B-L)_\omega + (B-L)_{\omega^*} =  2(r-\overline{r})$
%\end{equation}
The loop diagrams for $\omega\rightarrow \rho H^*$ are shown in Fig. \ref{baryo1} (b)-(d), which involve the exchange of fermions.
Since $\omega$ can also decay to two on--shell fermions, these loop diagrams have absorptive parts and also CP violation.

We evaluate Fig. \ref{baryo1} in a basis where the Majorana mass matrix $M_{\nu^c}$ of the $\nu^c$ fields is diagonal and real.  The contributions of
Fig. \ref{baryo1} (b)-(d) to $\epsilon_{B-L}$ are found to be
\begin{eqnarray}
\epsilon_{B-L} &=& -\frac{{\rm Br}}{\pi|\lambda v_R|^2} {\rm Im} \left[\lambda v_R \,{\rm Tr}\left\{Y^\dagger_{QL \omega^*} \, Y_{Q \nu^c \overline{\rho}}\,M_{\nu^c} \, F(M_\omega, M_\rho,M_{\nu^c_j})\, Y_{\nu^c L H}  \right. \right.\nonumber \\
&~& \left. \left.  - Y_{d^c \nu^c \omega} \,Y^\dagger_{d^c L \rho} \,\, Y_{\nu^c L H} \, M_{\nu^c}\, F(M_\rho, M_\omega,M_{\nu^c_j})\right.\right. \nonumber \\
&~& \left. \left.  - Y^\dagger_{d^c \nu^c \omega'} \, Y_{d^c \nu^c \omega}
\, F'(M_\omega, M_{\omega'},M_{j})  (\lambda'v_R)^* \right\}
 \right],
\label{asymb}
\end{eqnarray}
where the three terms are in order from Fig. \ref{baryo1} (b), (c) and (d).
Here we have defined the trilinear scalar vertices of Fig. \ref{baryo1} (a)  and (d) to have a coefficients $\lambda v_R$ and
$\lambda' v_R$ in the Lagrangian.
$Y_{Q \nu^c \overline{\rho}}$ is
the Yukawa coupling matrix corresponding to the coupling $Q \,\nu^c\, \overline{\rho}$, etc.
${\rm Br}$ stands for the branching ratio
${\rm Br}(\omega \rightarrow \rho H^*)$.  A factor of $2$ has been included here for the two $SU(2)_L$ final states in the decay.
The functions $F$ and $F'$ are defined as
\begin{eqnarray}
F(a,b,c) &=& {\rm ln}(1+a^2/c^2) + \Theta(1-c^2/b^2)\, (1-c^2/b^2), \nonumber \\
F'(a,b,c) &=& (1-c^2/a^2)/(1-b^2/a^2)\, \Theta(1-c^2/a^2)\,(1-c^2/a^2).
\end{eqnarray}
Here $\Theta$ stands for the step function, signalling additional ways of cutting the
diagram when $M_j < M_\rho$ or $M_j < M_\omega$ in Fig. \ref{baryo1}.
Fig. \ref{baryo1} (d) arises because in any realistic $SO(10)$ model there are at least two $\omega$ fields.  The heavier $\omega$
field is denoted as $\omega'$.
We have also
assumed that $M_\omega - M_\omega' \gg \Gamma_\omega$, so that there is no resonant enhancement for the decay.

To estimate  ${\rm Br} = {\rm Br}(\omega \rightarrow \rho H^*)$  appearing in Eqs. (\ref{asymb}),
let us assume that $\omega$ is the field $\omega$ from $10_H$ with Yukawa
couplings as given in Eq. (\ref{Yuk10}).  The partial widths for the decays $\Gamma_1(\omega \rightarrow \rho H^*)$ and $\Gamma_2(\rho
\rightarrow ff)$ are then given by
\begin{equation}
\Gamma_1(\omega \rightarrow \rho H^*) = \frac{|\lambda v_R|^2}{8 \pi M_\omega}\left(1-\frac{M_\rho^2}{M_\omega^2}  \right),~~~~~
\Gamma_2(\omega \rightarrow ff) = \frac{{\rm Tr}(h^\dagger h) }{4 \pi} M_\omega,
\label{partial}
\end{equation}
with ${\rm Br} = \Gamma_1/(\Gamma_1 + \Gamma_2)$.
For $M_\omega = 10^{16}$ GeV, $h_{33} = 0.6$ (corresponding to the top quark Yukawa
coupling at GUT scale) with other $h_{ij}$ negligible, and $\lambda v_R = (10^{14},\,10^{15},\,10^{16})$ GeV,
one gets ${\rm Br} = (1.4 \times 10^{-4},\, 1.4 \times 10^{-2},\,0.58)$.

The $(B-L)$ asymmetry $\epsilon_{B-L}$ of Eq. (\ref{asymb}) will result in a baryon to entropy ratio $Y_B$ given by
\begin{equation}
Y_B \equiv \frac{n_B-n_{\overline{B}}}{s} = \frac{\epsilon_{B-L}}{g_*} d~,
\label{eta}
\end{equation}
where $g_*=130$ is the total number of relativistic degrees of freedom at the epoch when these decays occur.
The factor $d$ in Eq. (\ref{eta}) is the dilution factor which takes into account back reactions that would
partially wash out the induced baryon asymmetry. %Simple analytic approximations for $d$
%are available suitable to the present setup.
Defining
$K = \left. \frac{\Gamma(\omega \rightarrow \rho H^*)} {2{\rm H}}\right|_{T=M_\omega}$, where ${\rm H}$ is the Hubble expansion rate,
${\rm H} = 1.66\, g_*^{1/2} \frac{T^2}{M_{\rm Pl}}$,
the dilution factor can be written as \cite{book}
$d\simeq 1 \,(K < 1)$, and $d\simeq \frac{0.3}{K ({\rm ln}\,K)^{0.6}}\, (K \gg 1)$.
For $M_\omega = 10^{15}$ GeV, and
$\lambda v_R = (10^{14},\,10^{15},\,10^{16})$ GeV, we find $K = (0.12,\,12.3,\,1230)$ and the corresponding
dilution factors to be $d= (1.0,\,1.4 \times 10^{-2},\, 7.5 \times 10^{-5})$, with ${\rm Br}= (1.3 \times 10^{-2},\,0.58,\,1.0)$.
%The baryon asymmetry generated in the decay $\eta \rightarrow \rho H^*$ is similar.

We now show how the GUT scale induced baryon asymmetry in $\omega \rightarrow \rho H^*$ decay
can consistently explain the observed value of $Y_B = (8.75 \pm 0.23) \times 10^{-11}$,
in a class of minimal $SO(10)$ models.  In these models a single $10_H$ and a single
$\overline{126}_H$ couple to fermions, as in  Eqs. (\ref{Yuk10})-(\ref{Yuk126}).  It has  been shown
that these models lead to large mixing angles for solar and atmospheric neutrino oscillations.  Furthermore, they
predict $\sin^2 2\theta_{13} \approx (0.085-0.095)$, both in the non--SUSY and the SUSY
versions \cite{babu}, which is consistent with recent results from Daya Bay and
other experiments \cite{theta13}.
To illustrate how realistic choice of parameters generate acceptable $Y_B$, we choose the
$\omega$ field to be almost entirely in the $10_H$.  We also choose $\lambda'v_R$ that appears
in Fig. \ref{baryo1} (d) to be small, so that the leading contribution to $\epsilon_{B-L}$ is
from Fig. \ref{baryo1} (c).  In this limit, we find
$\epsilon_{B-L} \approx \frac{2 \sqrt{3}}{\pi} \frac{|h_{33}f_3|^2}{|\lambda|}\{1+{\rm ln}(1+ M_\rho^2/M_{\nu_3^c}^2)\} \sin\phi$.
Here we have kept only the third family Yukawa couplings, and
defined $\phi = {\rm arg}\{h_{33}^2 f_3^2 \lambda + \frac{\pi}{2}\}$.  Choosing $h_{33} \simeq 0.6$ (the top quark Yukawa
coupling at the GUT scale), and $\lambda = 0.25$, $v_R = 10^{16}$ GeV, $f_3 = 10^{-2}$ (so that $f_3 v_R = 10^{14}$ GeV,
consistent with the light $\nu_\tau$ mass arising via the seesaw mechanism), $\phi = 0.12$, we find $\epsilon_{B-L} = 1.6 \times 1.9 \times 10^{-5}$.
If  $M_\omega = 10^{15}$ GeV, then ${\rm Br} = 0.96$, $K=197$ so that the dilution factor is $d=
5.6 \times 10^{-4}$.  This results in a net $Y_B= 8.2 \times 10^{-11}$, consistent with observations.  We emphasize the intimate connection
between $\epsilon_{B-L}$ and neutrino masses, since $Y_{QL\omega^*}^\dagger$, $Y_{Q\nu^c \overline{\rho}}$, etc present in $\epsilon_{B-L}$
are the ${\bf 126}$ couplings that determine the neutrino masses via the seesaw mechanism.

In conclusion, we have shown that all $d=7$ baryon number violating operators that lead to nucleon decay modes such as
$n \rightarrow e^- K^+, e^- \pi^+$ and $p \rightarrow \nu \pi^+$,
emerge naturally as effective low energy operators in a wide class of $SO(10)$ models.  In non--supersymmetric $SO(10)$ models with
an intermediate scale, we find the rates for these nucleon decay modes to be within reach of experiments. We have also
shown that the existence of these  $(B-L)$--violating interactions allows a new scenario for baryogenesis
where a $(B-L)$ asymmetry is generated in the decay of GUT mass particles which survives
to low temperatures unaffected by the sphaleron interactions. In minimal $SO(10)$
models which predict large neutrino mixing angles, including $\theta_{13}$, this new mechanism
can explain the observed baryon asymmetry of the universe.

The work of KSB is supported in part the US Department of Energy, Grant Numbers DE-FG02-04ER41306 and that of RNM  is supported
in part by the National Science Foundation Grant Number PHY-0968854.


\begin{thebibliography}{99}

\bibitem{weinberg}   S.~Weinberg, Phys.\ Rev.\ Lett.\  {\bf 43}, 1566 (1979);  F.~Wilczek and A.~Zee,
  %``Operator Analysis of Nucleon Decay,''
  Phys.\ Rev.\ Lett.\  {\bf 43}, 1571 (1979);L.~F.~Abbott and M.~B.~Wise,
  %``The Effective Hamiltonian For Nucleon Decay,''
  Phys.\ Rev.\ D {\bf 22}, 2208 (1980).


  \bibitem{weinberg2} S.~Weinberg,
  %``Varieties of Baryon and Lepton Nonconservation,''
  Phys.\ Rev.\ D {\bf 22}, 1694 (1980); H.~A.~Weldon and A.~Zee,
  %``Operator Analysis Of New Physics,''
  Nucl.\ Phys.\ B {\bf 173}, 269 (1980).
  %%CITATION = NUPHA,B173,269;%%

   \bibitem{kuzmin} V.~A.~Kuzmin, V.~A.~Rubakov and M.~E.~Shaposhnikov,
  %``On the Anomalous Electroweak Baryon Number Nonconservation in the Early Universe,''
  Phys.\ Lett.\ B {\bf 155}, 36 (1985).


  \bibitem{babu}  K.~S.~Babu, R.~N.~Mohapatra,
  %``Predictive Neutrino Spectrum In  SO(10) Grand Unification,''
  Phys.\ Rev.\ Lett.\  {\bf 70}, 2845 (1993);
   B.~Bajc, G.~Senjanovic and F.~Vissani,
  %``How neutrino and charged fermion masses are connected within minimal supersymmetric SO(10),''
  hep-ph/0110310;
  %``b - tau unification and large atmospheric mixing: A Case for noncanonical seesaw,''
  Phys.\ Rev.\ Lett.\  {\bf 90}, 051802 (2003);
   T.~Fukuyama and N.~Okada,
  %``Neutrino oscillation data versus minimal supersymmetric SO(10) model,''
  JHEP {\bf 0211}, 011 (2002);
   %%CITATION = HEP-PH/0205066;%%
  H.~S. Goh, R.~N. Mohapatra, S.~P. Ng, Phys. Lett. {\bf B570}, 215 (2003);
     K.S. Babu, C. Macesanu, Phys. Rev. {\bf D72}, 115003 (2005);
	  S. Bertolini, T. Schwetz, M. Malinsky, Phys. Rev. {\bf D73}, 115012 (2006);
	    A.~S.~Joshipura, K.~M.~Patel,
  %``Viability of the exact tri-bimaximal mixing at M_{GUT} in SO(10),''
  arXiv:1105.5943 [hep-ph].



\bibitem{theta13} F.~P.~An {\it et al.}  [DAYA-BAY Collaboration],
  %``Observation of electron-antineutrino disappearance at Daya Bay,''
  arXiv:1203.1669 [hep-ex].
 K.~Abe {\it et al.}  [T2K Collaboration],
  %``Indication of Electron Neutrino Appearance from an Accelerator-produced
  %Off-axis Muon Neutrino Beam,''
  Phys. Rev. Lett. {\bf 107}, 041801 (2011);
   P. Adamson {\it et al.} [MINOS Collaboration],
  Phys.\ Rev.\ Lett.\  {\bf 107}, 181802 (2011);
	%MINOS Collaboration (2011),
H. De. Kerrect [Double CHOOZ Collaboration], talk
	at the LowNu conference in Seoul, Korea (2011).% {\tt http://workshop.kias.re.kr/lownu11/?Program}.





\bibitem{bm2012}For further details see:
 K.~S.~Babu and R.~N.~Mohapatra,
  %``B-L Violating Nucleon Decay and GUT Scale Baryogenesis in SO(10),''
    arXiv:1203.5544 [hep-ph]
  (to be submitted to Phys. Rev. D).


\bibitem{chang}   R.~N.~Mohapatra and M.~K.~Parida,
  %``Threshold effects on the mass scale predictions in SO(10) models and solar neutrino puzzle,''
  Phys.\ Rev.\ D {\bf 47}, 264 (1993);
  S.~Bertolini, L.~Di Luzio and M.~Malinsky,
  %``Intermediate mass scales in the non-supersymmetric SO(10) grand unification: A Reappraisal,''
  Phys.\ Rev.\ D {\bf 80}, 015013 (2009).



  \bibitem{aulakh1}
  C.~S.~Aulakh and A.~Girdhar,
  %``SO(10) MSGUT: Spectra, couplings and threshold effects,''
  Nucl.\ Phys.\ B {\bf 711}, 275 (2005);
    %%CITATION = HEP-PH/0405074;%%
 C.~S.~Aulakh and S.~K.~Garg,
  %``The New Minimal Supersymmetric GUT : Spectra, RG analysis and fitting formulae,''
  hep-ph/0612021.
  %%CITATION = HEP-PH/0612021;%%



   \bibitem{vissani}
   F.~Vissani,
  %``(B+L) conserving nucleon decays in supersymmetric models,''
  Phys.\ Rev.\ D {\bf 52}, 4245 (1995).



  \bibitem{maekawa}  S.~Enomoto and N.~Maekawa,
  %``Baryogenesis by B - L generation due to superheavy particle decay,''
  Phys.\ Rev.\ D {\bf 84}, 096007 (2011);
   P.~-H.~Gu and U.~Sarkar,
  %``Common origin of baryon asymmetry and proton decay,''
  arXiv:1110.4581 [hep-ph].


  \bibitem{book}
 {\it The Early Universe}, by E. Kolb and M. Turner, (Frontiers in Physics) (1986).







\end{thebibliography}
\end{document}